\documentclass[preprint, 12pt, a4paper]{elsarticle}

\usepackage{comment}
\usepackage{amssymb}
\usepackage{hyperref}
\usepackage{listings}
\usepackage{xcolor}
\usepackage{orcidlink}
\journal{SoftwareX}

\begin{document}
\renewcommand{\labelenumii}{\arabic{enumi}.\arabic{enumii}}

\begin{frontmatter}

\title{SGN: A python framework for stream-processing pipelines}


\author[label3,label1,label2]{Yun-Jing Huang \orcidlink{0000-0002-2952-8429}}
\ead{yun-jing.huang@ligo.org}


\author[label3]{Olivia Godwin \orcidlink{0000-0002-7489-4751}}
\author[label1,label2,label4,label5]{Chad Hanna \orcidlink{0000-0002-0965-7493}}
\author[label1,label2]{James Kennington \orcidlink{0000-0002-6899-3833}}
\author[label3]{Jameson Rollins \orcidlink{0000-0002-9388-2799}}

\author[label3]{Max Melching \orcidlink{0009-0001-4899-9955}}
\author[label6]{Nathanael E Sovitzky}
\author[label6]{Aaron Viets \orcidlink{0000-0002-4241-1428}}
\author[label7]{Madeline Wade \orcidlink{0000-0002-5703-4469}}
\author[label8]{Zach Yarbrough \orcidlink{0000-0002-9825-1136}}

\author[label9]{Yu-Kuang Chu \orcidlink{0000-0002-8661-4120}}
\author[label7]{William Wyatt Phillips}
\author[label10]{Surabhi Sachdev \orcidlink{0000-0002-0525-2317}}
\author[label11]{Rhiannon Udall \orcidlink{0000-0001-6877-3278}}

\address[label3]{LIGO Laboratory, California Institute of Technology, Pasadena, California 91125, USA}
\address[label1]{Department of Physics, The Pennsylvania State University, University Park, PA 16802, USA}
\address[label2]{Institute for Gravitation and the Cosmos, The Pennsylvania State University, University Park, PA 16802, USA}

\address[label4]{Department of Astronomy and Astrophysics, The Pennsylvania State University, University Park, PA 16802, USA}
\address[label5]{Institute for Computational and Data Sciences, The Pennsylvania State University, University Park, PA 16802, USA}
\address[label6]{Concordia University Wisconsin, Mequon, WI 53097, USA}
\address[label7]{Department of Physics, Hayes Hall, Kenyon College, Gambier, Ohio 43022, USA}
\address[label8]{Department of Physics and Astronomy, Louisiana State University, Baton Rouge, LA 70803, USA}
\address[label9]{Leonard E.\ Parker Center for Gravitation, Cosmology, and Astrophysics, University of Wisconsin-Milwaukee, Milwaukee, WI 53201, USA}
\address[label10]{School of Physics, Georgia Institute of Technology, Atlanta, GA 30332, USA}
\address[label11]{Department of Physics \& Astronomy, University of British Columbia, Vancouver, BC V6T 1Z1, Canada}

\begin{abstract}
We present the Stream Graph Navigator (SGN), a lightweight Python framework for building streaming data applications. In SGN, stream-processing pipelines are built by connecting computational components into directed acyclic graphs that run within an event loop. The time-series extension of the SGN library, SGN-TS, introduces signal processing methods to handle time series data. Together, SGN and SGN-TS provide the foundation for SGNL, a matched-filtering gravitational-wave search pipeline, and are being adopted by multiple projects across the low-latency gravitational-wave data analysis infrastructure as an extensible and maintainable framework for future gravitational-wave observations.


\end{abstract}


\end{frontmatter}


\section*{Metadata}
\label{}

\begin{table}[!h]
\begin{tabular}{|l|p{6.5cm}|p{6.5cm}|}
\hline
\textbf{Nr.} & \textbf{Code metadata description} & \textbf{Metadata} \\
\hline
C1 & Current code version & sgn: 0.11.0, sgn-ts: 0.9.2\\
\hline
C2 & Permanent link to code/repository used for this code version & \url{https://git.ligo.org/greg/sgn/}, \url{https://git.ligo.org/greg/sgn-ts/}
\\
\hline
C3  & Permanent link to Reproducible Capsule & \url{https://pypi.org/project/sgn/}, \url{https://pypi.org/project/sgn-ts/}\\
\hline
C4 & Legal Code License   & Mozilla Public License 2.0 \\
\hline
C5 & Code versioning system used & git \\
\hline
C6 & Software code languages, tools, and services used & Python \\
\hline
C7 & Compilation requirements, operating environments \& dependencies & sgn: Python $>=$ 3.10; sgn-ts: Python $>=$ 3.10, Numpy, Scipy \\
\hline
C8 & If available Link to developer documentation/manual & \url{https://greg.docs.ligo.org/sgn/}, \url{https://greg.docs.ligo.org/sgn-ts/} \\
\hline
C9 & Support email for questions & yun-jing.huang@ligo.org \\
\hline
\end{tabular}
\caption{Code metadata (mandatory)}
\label{codeMetadata} 
\end{table}

\section{Motivation and significance}


Gravitational waves (GWs) are perturbations in spacetime predicted by general relativity \cite{Einstein:1916cc,Einstein:1915ca}. They can be produced by accelerated masses, such as two compact objects, black holes or neutron stars, orbiting and merging together. As the binary system evolves, it loses energy to GWs, and the waves propagate across the Universe, encoding information about the properties of the source. Since the amplitudes of GWs are very weak when they arrive at Earth, very precise detectors are needed to measure these tiny changes in spacetime. Current ground-based GW detectors, including LIGO \cite{ligo}, Virgo \cite{virgo}, and KAGRA \cite{kagra}, are kilometer-scale laser interferometers in which passing GWs slightly change the relative lengths of the arms, altering how the laser light recombines at the output. This detector output is then converted, through calibration, into a time series of strain data, which describes the fractional change in arm length caused by the passing wave. Sophisticated computational techniques are then applied to extract weak GW signals from the noisy strain data. Advancements in the development of interferometers and data analysis infrastructure led to the first direct observation of GWs from a binary black hole merger in 2015 \cite{gw150914}. Since then, the growing catalog of GW events reported by the LIGO–Virgo–KAGRA collaboration (LVK) \cite{gwtc-1,gwtc-2.1,gwtc-3,gwtc-4,gwtc-5}, together with the regular release of low-latency public alerts during the recently concluded fourth observing run \cite{gracedb}, shows that GW detections have become routine. Maintaining this capability requires robust and sustainable data analysis infrastructure as new detectors begin operation \cite{kagra,indigo1,indigo2,lisa} and the detection rate increases. A fast and reliable low-latency detection framework is also critical for multimessenger follow-up and maximizing scientific return.

The current ground-based interferometers are sensitive to GW signals in the audio-frequency band, and their output is therefore represented as audio-frequency strain time series. These time series are analyzed continuously as new data arrive to enable rapid event alerts. This made the GStreamer multimedia framework \cite{gstreamer} a natural foundation for GstLAL, a matched-filtering based compact-binary-coalescence (CBC) search pipeline used to identify GW signals in streaming detector data \cite{messick2017,CANNON2021,ewing2024,tsukada}. GStreamer provided stream-processing tools that could be adapted to the needs of low-latency GW data analysis. GstLAL has participated in every observing run and has routinely identified GW candidates, including the first binary neutron star merger detected in low latency \cite{gw170817}. Its stream-processing tools also support applications beyond the CBC search, including \texttt{gstlal-calibration} \cite{viets2018,wade2023,wade2025}, which converts detector output into calibrated strain data, and the Stream-based Noise Acquisition and eXtraction (SNAX) pipeline \cite{godwin2020low}, which identifies non-Gaussian transient noise in the data. Although these pipelines have been successfully deployed during observing runs, the underlying C-based internals are difficult to maintain and extend. Modernizing the GstLAL and GStreamer-based framework is therefore important for the long-term sustainability of the GW data analysis ecosystem.

The Stream Graph Navigator (SGN) library was developed to provide a more sustainable software framework for streaming applications in GW data analysis. SGN is a lightweight Python framework that draws on the core streaming concepts of GStreamer. Processing steps are represented by elements, and data are passed between elements through pad interfaces. A full SGN pipeline is represented as a directed acyclic graph (DAG), and the event loop manages the execution of each processing step. Its time-series extension, SGN-TS, builds on this framework by adding methods for time-series data handling and signal processing. By using SGN and SGN-TS as a common software base, streaming pipelines across the GW analysis stack can be extended and adapted to multiple applications over time.

\section{Software description}


\subsection{Software architecture}

\subsubsection{Core framework: SGN}

\begin{figure}
    \centering
    \includegraphics[width=1\linewidth]{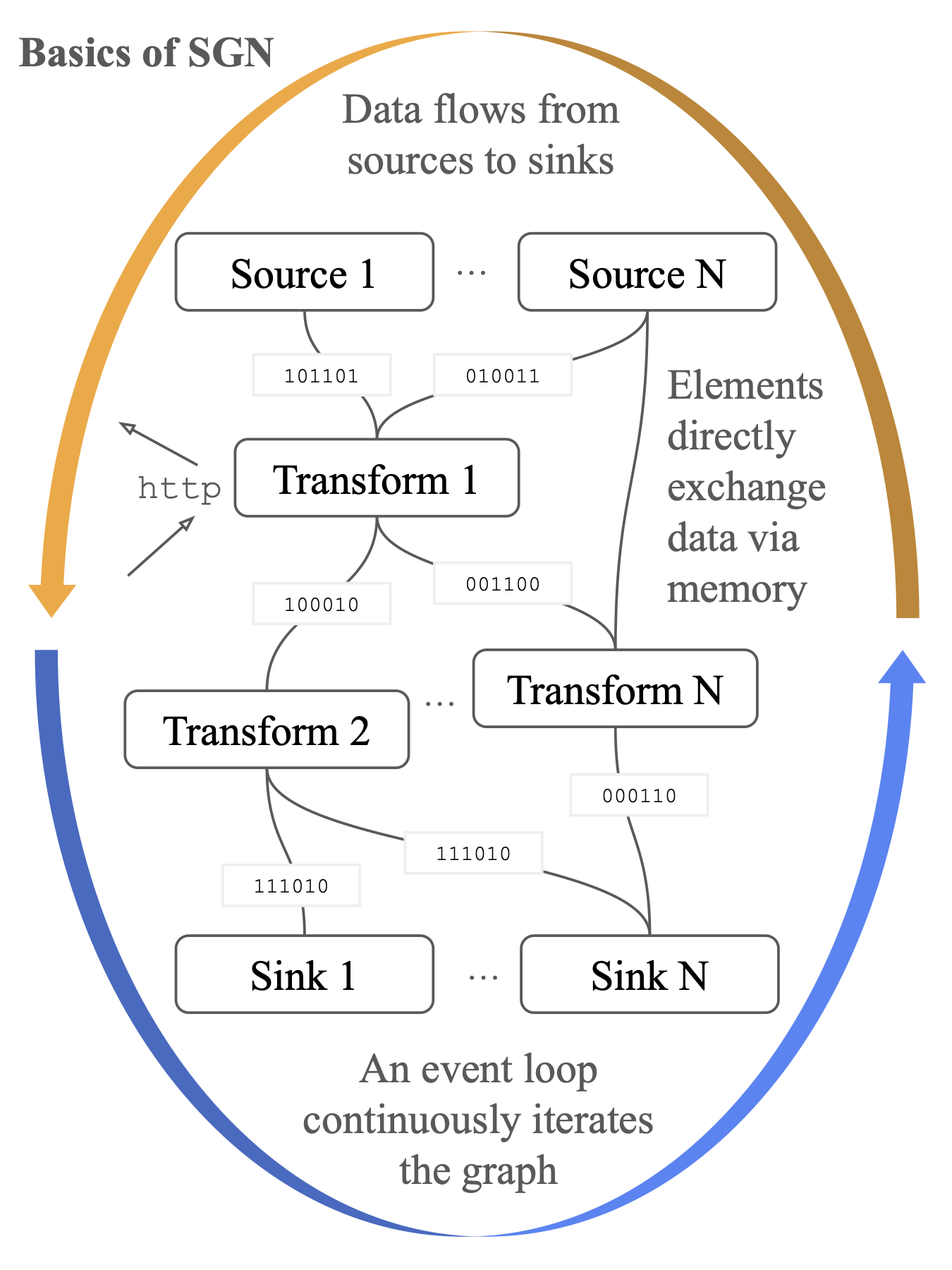}
    \caption{Diagram of the DAG model in the SGN framework. Multiple source, transform, and sink elements can be connected through their pad interfaces. Frames flow through the pipeline and carry data. The data are exchanged between the pads on the element boundaries through memory. The DAG is executed in a continuous event loop. The additional feature, HTTP control, allows external operators to update the element configuration while the pipeline is running.}
    \label{fig:sgn}
\end{figure}

SGN is a data streaming framework that is built on a DAG task execution model. It is a lightweight Python package that depends only on the Python standard library. At its core, the framework is organized around four key concepts, ascending in scope: frames, pads, elements, and pipeline. Figure \ref{fig:sgn} shows an overview of how elements and pads are connected to build the DAG framework in SGN.

\paragraph{Frames} A frame is the data unit that flows through the pipeline. In addition to the data payload, it carries the information needed to handle that data during execution, including metadata, a gap flag for missing or invalid data, an end-of-stream (EOS) flag, and a type field.

\paragraph{Pads} Pads are nodes in the DAG model. They serve as the connection points that link elements into a pipeline and are implemented as callable dataclasses, so each pad is executed once during every iteration of the DAG. Pads belong to elements and can be divided into three types with respect to the element: source, sink, and internal. A source pad emits a frame object from an element, whereas a sink pad consumes a frame into the element. An internal pad comes with each element, and lies before source pads and after sink pads. It is the place where data processing methods are performed after an element has consumed frames from all the sink pads and before emitting frames through all the source pads.

\paragraph{Elements} Elements are the basic processing components that form the pipeline. They are dataclasses that define computations in the pipeline and maintain any internal state needed to carry them out, while pads determine how frames enter and leave the element. Elements can be classified as sources, transforms, and sinks. Source elements define entry points to the pipeline, and emit frames from their source pads to introduce new frames into the rest of the pipeline. Transform elements define operations in the pipeline that consume frames through their sink pads, modify them, and emit new frames through their source pads to downstream elements. Sink elements are the end points of the pipeline. They consume frames through their sink pads and the final output may be written to disk or transmitted out of the pipeline.

\paragraph{Pipeline} A pipeline is formed by connecting a set of elements through their pads. It is a DAG that is executed repeatedly in an event loop. Elements are linked together by connecting their sink pads to the source pads of upstream elements.
After linking elements, the application uses a topological sort to construct the dependency graph. 
This graph structure determines how data move through the pipeline and ensures that upstream computation completes before downstream pads consume its output. An event loop executes the DAG repeatedly until EOS signals are received by all the sink pads. During each iteration, every pad is called once, so each source pad emits one frame and each sink pad consumes one frame. This execution model makes the behavior of the pipeline explicit and deterministic.

These four abstractions are the basic components of the framework. In addition, there are advanced features in the SGN library including composition, parallelization through multiprocessing, and context managers to allow graceful shutdown and HTTP control for the pipeline.

\paragraph{Composition} Elements can be grouped together to act as a single composite element within the pipeline. The internal elements remain connected through their pads, while only the boundary pads are exposed. This allows common connection patterns to be reused across different pipelines.

\paragraph{Parallel execution} The parallelize extension allows elements to run in separate subprocesses. Data move between the main process and the subprocess through a queue, which makes it possible to offload work without changing the DAG structure of the pipeline.

\paragraph{Signal control} The signal control feature allows the pipeline to shut down gracefully. When the pipeline runs under the \texttt{SignalEOS} context manager, a \texttt{SIGINT} or \texttt{SIGTERM} signal triggers an orderly shutdown that propagates the EOS signal throughout the pipeline. This is especially useful in real-time applications, where pipelines may run indefinitely and need to be stopped or restarted to apply updates.

\paragraph{HTTP control} The \texttt{HTTP} control feature allows an external operator to update the pipeline while it is running. Under this context manager, the pipeline will launch a small \texttt{Bottle} web service in a separate thread, which exposes \texttt{GET} and \texttt{POST} endpoints to the elements. An operator can query or update element state through \texttt{HTTP} without terminating and restarting the pipeline. When an update is posted to an element endpoint, the change takes effect in the next iteration of the DAG execution loop. This is useful for real-time GW search pipelines, where taking down and relaunching a pipeline risks missing critical events.

\subsubsection{Time-series extension: SGN-TS}
The SGN-TS package is a time-series extension of the SGN core framework. It adds the concept of time and includes methods to handle time series data, including precise timing, synchronization across multiple data streams, slicing and merging data buffers, and sliding window computation. In addition to the Python standard library, SGN-TS also depends on NumPy, Scipy, and has an optional dependency on PyTorch. The time series data can be represented as NumPy arrays or PyTorch tensors. This allows computation to optionally use modern methods such as GPU acceleration and machine-learning tools through PyTorch.

\paragraph{Offset system}
An accurate timing system is critical for GW data analysis. Matched filtering pipelines require comparing a template across the data and finding correlations through the data. Inaccurate timing might lower the recovered signal-to-noise ratio and lead to missed events. GW data lie in the audio frequency range and are mostly power-of-two sample rates, and precise synchronization between different sample rates is crucial for signal processing. The offset system provides an integer timing scheme where the sample points of data with power-of-two rates will lie exactly on offset boundaries. This avoids accumulating rounding errors that can arise when using a nanosecond timing system, and ensures that samples at different rates remain aligned. The current implementation requires time-series data passed through the pipeline to have power-of-two sample rates and rejects non-power-of-two sample rates. Supporting additional sample rates would require extending the offset system and is left for future work.

\paragraph{Frames and Buffers} In a streaming setting, time-series data may include both valid segments and gaps, so SGN-TS needs a representation that can describe both within the same frame. The data that flow through pipelines built with SGN-TS are carried by \texttt{TSFrames}, which subclass the Frame object in SGN. A \texttt{TSFrame} contains a list of \texttt{SeriesBuffer} objects. Each \texttt{SeriesBuffer} represents a block of time-series data through its starting offset, duration, sample rate, and gap flag, allowing it to describe either valid data or a gap with no valid samples.

\paragraph{Audioadapter} The \texttt{audioadapter} is a queuing system that manages the bookkeeping of streaming data. It stores incoming buffers in a queue and keeps track of their ordering in time, requiring consecutive buffers to remain continuous. Additional methods are defined to allow users to copy certain segments of data from the queue or examine gap segments. Signal processing methods often require processes to perform computation on data over a fixed stride or with a certain overlap from earlier or later samples, such as overlap-save methods or convolution. The \texttt{audioadapter} provides this buffering and slicing layer so that elements can prepare the data they need without re-implementing the same queuing logic.

\paragraph{Array backend} The array backend abstraction allows elements to be written in a backend-agnostic way. SGN-TS currently supports NumPy and PyTorch, and can be extended to other numerical libraries. PyTorch support allows pipelines to use GPU computing without explicitly maintaining separate code versions for different hardware.

\subsection{Software functionalities}

\subsubsection{SGN}

\paragraph{Base classes} The SGN framework provides base classes that users can extend for new applications. The \texttt{SourceElement}, \texttt{TransformElement}, and \texttt{SinkElement} classes define the main patterns for building custom elements. In practice, users implement the pad methods that carry out the element logic: \texttt{new()} for source pads, \texttt{pull()} for sink pads, and \texttt{internal()} for the internal pad. Once these methods are defined and the elements are connected, the framework handles dependency tracking and execution of the graph.

\paragraph{Standard elements} The SGN package also includes a small set of ready-made elements for simple streaming tasks, such as \texttt{IterSource} and \texttt{CollectSink} for iterating through data and collecting output, and \texttt{NullSource} and \texttt{NullSink} for testing simple pipelines.

\subsubsection{SGN-TS}

\paragraph{Base classes} The SGN-TS framework also provides base classes for building custom elements: \texttt{TSSource}, \texttt{TSTransform}, and \texttt{TSSink}. The transform and sink base classes provide automatic multi-channel alignment of incoming frames across sink pads and check the continuity of the incoming data stream, ensuring that no discontinuities or overlaps occur.

\paragraph{Adapter configuration} Elements that require strided, overlapping, or sliding-window computation can be configured with an optional \texttt{AdapterConfig}, so they do not need to re-implement the same streaming logic themselves. When \texttt{AdapterConfig} is used, each sink pad is given an audioadapter instance to handle the buffering needed for these operations. \texttt{AdapterConfig} also supports alignment to specific time boundaries and handling of gaps.

\paragraph{Signal processing library} The SGN-TS package comes with existing signal processing elements, such as resampling, correlate, matrix multiplication, and gating. These elements can be connected directly into a pipeline or used as starting points for custom implementations.


\section{Illustrative examples}



\subsection{SGN: Simple counting pipeline}
\begin{figure}
    \centering
    \includegraphics[width=0.7\linewidth]{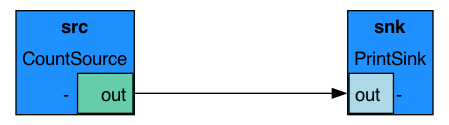}
    \caption{Graph of the simple counting pipeline. There is one source element and one sink element, and they are connected through their source and sink pads.}
    \label{fig:count}
\end{figure}

The following example shows a simple counting pipeline, where a source element generates an integer that increases by one in each loop iteration, and a sink element prints the value it receives. Figure \ref{fig:count} shows the corresponding graph, where pads with matching names are linked automatically through the \texttt{connect()} method. In each iteration, the source element produces a new \texttt{Frame} through its \texttt{new()} method, and the sink element consumes that frame through its \texttt{pull()} method. The EOS condition is defined by the source element as the condition when the number reaches the limit.

\begin{lstlisting}[language=Python, caption={Simple counting pipeline.}, label={code:count}]
from dataclasses import dataclass
from sgn import SourceElement, SinkElement, SourcePad, SinkPad, Frame, Pipeline

@dataclass
class CountSource(SourceElement):
    """Produces frames counting from 1 to `limit`."""
    limit: int = 3

    def __post_init__(self):
        super().__post_init__()
        self.count = 0

    def new(self, pad: SourcePad) -> Frame:
        self.count += 1
        return Frame(
            data=self.count,
            EOS=self.count >= self.limit,
        )

@dataclass
class PrintSink(SinkElement):
    """Prints each frame's data."""
    def pull(self, pad: SinkPad, frame: Frame) -> None:
        print(f"{pad.pad_name}: {frame.data}")
        if frame.EOS:
            self.mark_eos(pad)
            return


src = CountSource(name="src", source_pad_names=["out"])
snk = PrintSink(name="snk", sink_pad_names=["out"])

p = Pipeline()
p.connect(src, snk)
p.run()
\end{lstlisting}

The output of the pipeline is shown as follows. The pipeline goes through three iterations, and in each iteration the \texttt{PrintSink} element produces the a number, which is incremented by one after each iteration.
\begin{lstlisting}[language=bash, caption={Output of simple counting pipeline.}, label={code:count_out}]
out: 1
out: 2
out: 3
\end{lstlisting}

\subsection{SGN-TS: Downsampling a sine wave}
\begin{figure}
    \centering
    \includegraphics[width=1\linewidth]{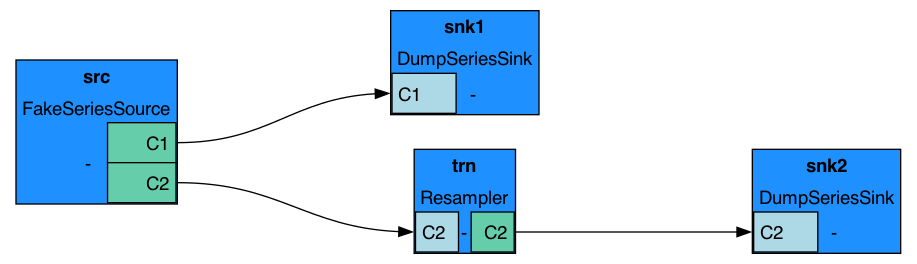}
    \caption{Pipeline graph for downsampling example.}
    \label{fig:downsample}
\end{figure}

This example illustrates how SGN-TS handles streaming time-series data with operations that require overlap for context. A simple pipeline generates a sine wave, downsamples it to a lower rate by filtering the data with a downsampling kernel, and writes the original frames and downsampled frames into files for comparison. The pipeline graph is shown in Figure \ref{fig:downsample}, and the pipeline construction is demonstrated in the following example code.

\begin{lstlisting}[language=Python, caption={Downsampling a sine wave using SGN-TS.}]
from sgn import Pipeline
from sgnts.sources import FakeSeriesSource
from sgnts.transforms import Resampler
from sgnts.sinks import DumpSeriesSink

p = Pipeline()

src = FakeSeriesSource(
        name='src',
        source_pad_names=('C1','C2'),
        rate=128,
        signal_type='sin',
        start=0,
        end=3,
        )

trn = Resampler(
        name='trn',
        sink_pad_names=('C2',),
        source_pad_names=('C2',),
        inrate=128,
        outrate=64,
        )

snk1 = DumpSeriesSink(
        name='snk1',
        sink_pad_names=('C1',),
        fname='source_output.txt',
        )

snk2 = DumpSeriesSink(
        name='snk2',
        sink_pad_names=('C2',),
        fname='downsampled.txt',
        )

p.connect(src,trn).connect(src,snk1).connect(trn,snk2)
p.run()
\end{lstlisting}

\begin{figure}
    \centering
    \includegraphics[width=\linewidth]{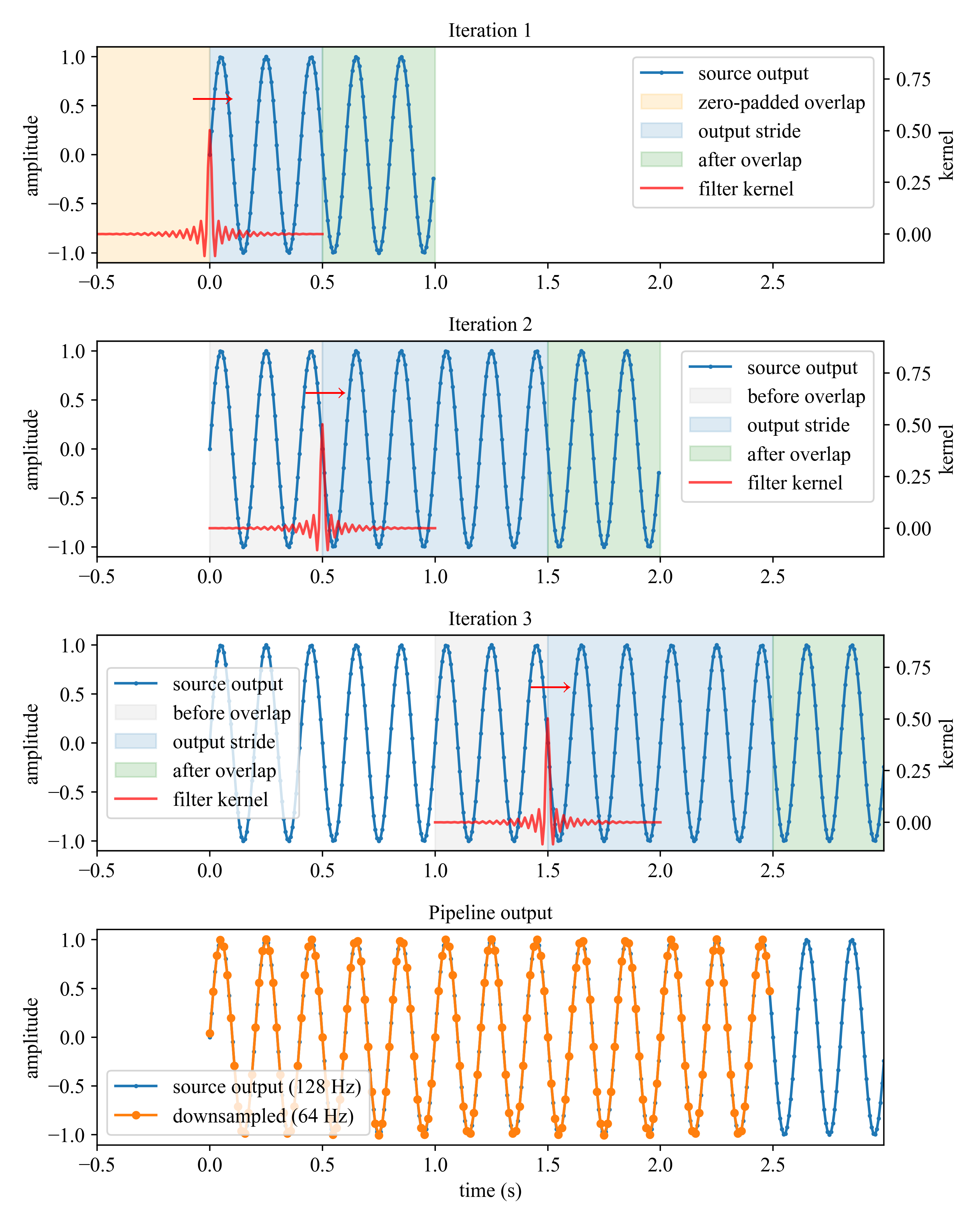}
    \caption{The streaming behavior in the downsampling example pipeline. The top three panels show a sine wave at sample rate 128 Hz being downsampled by filtering a sinc-windowed sinc function (red) across the streaming sine wave data frame by frame. Overlaps of earlier (grey shaded region) and later (green shaded region) input samples need to be padded as context to perform the continuous filtering. The bottom panel shows the result of the downsampling, where the 128 Hz sine wave (blue), is downsampled to 64 Hz (orange). The output lags the input by the amount of data in the green overlap region.}
    \label{fig:downsample_output}
\end{figure}

The \texttt{FakeSeriesSource} element generates a sine wave and outputs it through two source pads. One output is connected to the \texttt{Resampler} transform element, while the other is connected directly to a \texttt{DumpSeriesSink} element. The output of the \texttt{Resampler} is then sent to a second sink element, which writes the downsampled data to a file. When the pipeline runs, the \texttt{Resampler} downsamples the sine wave frame by frame, illustrating the streaming behavior of the pipeline. The top three panels in Figure \ref{fig:downsample_output} show how overlap is handled when downsampling a streaming sine wave over three iterations. The downsampling kernel, shown in red, is a sinc-windowed sinc function that sweeps across the shaded regions to filter the incoming data. In each iteration, the target output stride must be padded on both sides by half the kernel length using surrounding samples, so the filtering step has the context it needs to produce results. The bottom panel shows both the original input data and the downsampled output. The downsampled output lags the input by the amount of future context required by the filter.

\section{Impact}

One of the direct impacts of SGN is that it provides the software foundation for SGNL, a modernized version of the GstLAL CBC search pipeline, where the GStreamer-based framework has been replaced by SGN \cite{Huang2025_sgnl}. This is an example of SGN extending beyond a simple framework design and shows that it already supports a real low-latency GW search pipeline with production-scale requirements. In LVK mock data challenges, SGNL achieved GW search sensitivity comparable to GstLAL while reducing median latency from 9.0 seconds to 4.7 seconds \cite{Huang2025_sgnl}. This demonstrates that the GStreamer layer of GstLAL can be replaced by SGN while maintaining scientific performance. PyTorch integration further enabled a 169$\times$ speedup on a GPU compared to GstLAL running on a single CPU core \cite{huang2025}. These results show that SGN can support real-time GW analysis at the operation scale and take advantage of modern hardware.

SGN also changes how several GW analysis pipelines are developed. The previous GStreamer-based stack required developers to work within a C-based plugin framework, which required familiarity with its internal scheduling and element machinery, and made development difficult. SGN is written entirely in Python, so new elements and applications can be developed without significant framework overhead. Being written in Python also helps in the development of the pipeline execution layer. In large streaming applications, much of the complexity comes from the framework that orchestrates data propagation through the pipeline rather than from the signal processing algorithms themselves. Real-time constraints add to this, since graph execution and data movement must be managed so that data flow continuously and alerts are still produced reliably when unexpected interruptions or delays occur. In GStreamer, diagnosing problems in this layer is opaque and often required working through large debug logs. In SGN, the graph is a DAG, and the execution order is fully deterministic, which makes SGN easier to develop and debug than the previous GStreamer-based stack.

Another impact is the extension to modern hardware and software libraries. SGN-TS can use either NumPy or PyTorch backends, and the pipeline logic does not need to change when moving between CPUs and GPUs. This is especially relevant for CBC searches, where matched filtering is the dominant computational cost. With PyTorch support, GPU acceleration can be added directly to streaming pipelines without changing the overall pipeline design. It also makes it easier to connect these pipelines to machine-learning tools and other array-based numerical libraries. This makes SGN a more practical framework for exploring GPU-accelerated filtering and machine-learning-based inference methods in GW analysis.

The impact of SGN also extends beyond the SGNL inspiral search pipeline. SGN and its extension libraries are already being adopted in other parts of the LVK low-latency ecosystem that previously relied on GstLAL’s GStreamer-based tools. Calibration is an example, with the low-latency strain production pipeline now moving from \texttt{gstlal-calibration} to SGN-CAL. Detector characterization workflows such as SNAX, which depends on GstLAL stream processing tools, are also being reworked with SGN as SGNAX \cite{sgnax}, extending the SGN framework into data quality analysis. This pushes SGN upstream into other core parts of the low-latency stack.

More broadly, the SGN software family provides a general framework for data-processing pipelines. Although it was motivated by GW data analysis pipelines, SGN itself is a general task-graph execution model that can be subclassed and extended for other domains. SGN-TS extends it to time-series data and can itself serve as a base for signal-processing applications beyond GW astronomy. New applications in other domains can be built on top of this infrastructure without having to reimplement the core pipeline machinery.

\section{Conclusions}
The SGN library provides a general stream-processing framework for building streaming data applications. Together with its time-series extension, SGN-TS, it provides an extensible foundation for signal-processing applications. Multiple GW pipelines are adopting SGN to modernize their software base, and SGNL has already shown that the legacy framework can be replaced while still supporting production-level analysis. Although SGN was developed for GW analysis, the framework can be applied more broadly to other streaming and signal-processing applications.

\section*{Acknowledgements}
This material is based upon work supported by NSF's LIGO Laboratory which is a major facility fully funded by the National Science Foundation.
This research has made use of data, software and/or web tools obtained from the
Gravitational Wave Open Science Center (https://www.gw-openscience.org/ ), a
service of LIGO Laboratory, the LIGO Scientific Collaboration and the Virgo
Collaboration.  We especially made heavy use of the LVK Algorithm
Library~\cite{lalsuite}. LIGO Laboratory and Advanced LIGO are funded
by the United States National Science Foundation (NSF) as well as the Science
and Technology Facilities Council (STFC) of the United Kingdom, the
Max-Planck-Society (MPS), and the State of Niedersachsen/Germany for support of
the construction of Advanced LIGO and construction and operation of the GEO600
detector. Additional support for Advanced LIGO was provided by the Australian
Research Council.  Virgo is funded, through the European Gravitational
Observatory (EGO), by the French Centre National de Recherche Scientifique
(CNRS), the Italian Istituto Nazionale di Fisica Nucleare (INFN) and the Dutch
Nikhef, with contributions by institutions from Belgium, Germany, Greece,
Hungary, Ireland, Japan, Monaco, Poland, Portugal, Spain.

The authors are grateful for computational resources provided by the 
the Pennsylvania State University's Institute for Computational and Data
Sciences gravitational-wave cluster, and the LIGO Lab cluster at the LIGO Laboratory, and supported by 
the National Science Foundation awards 
OAC-2103662, PHY-2308881, OAC-2201445, OAC-2018299, PHY-0757058, PHY-0823459, PHY-2513124, and PHY-2309200.
CH Acknowledges generous support from the Eberly College of Science, the 
Department of Physics, the Institute for Gravitation and the Cosmos, and the 
Institute for Computational and Data Sciences.

\label{}



\bibliographystyle{elsarticle-num} 
\bibliography{references}

@article{ewing2024,
  title = "{Performance of the low-latency GstLAL inspiral search towards LIGO, Virgo, and KAGRA's fourth observing run}",
  author = {Ewing, Becca and others},
  journal = {Phys. Rev. D},
  volume = {109},
  issue = {4},
  pages = {042008},
  numpages = {18},
  year = {2024},
  month = {Feb},
  publisher = {American Physical Society},
  doi = {10.1103/PhysRevD.109.042008},
  url = {https://link.aps.org/doi/10.1103/PhysRevD.109.042008}
}

@article{messick2017,
  title = {Analysis framework for the prompt discovery of compact binary mergers in gravitational-wave data},
  author = {Messick, Cody and others},
  journal = {Phys. Rev. D},
  volume = {95},
  issue = {4},
  pages = {042001},
  numpages = {15},
  year = {2017},
  month = {Feb},
  publisher = {American Physical Society},
  doi = {10.1103/PhysRevD.95.042001},
  url = {https://link.aps.org/doi/10.1103/PhysRevD.95.042001}
}

@article{gw150914,
  title = {Observation of Gravitational Waves from a Binary Black Hole Merger},
  author = {Abbott, B. P. and others},
  collaboration = {LIGO Scientific Collaboration and Virgo Collaboration},
  journal = {Phys. Rev. Lett.},
  volume = {116},
  issue = {6},
  pages = {061102},
  numpages = {16},
  year = {2016},
  month = {Feb},
  publisher = {American Physical Society},
  doi = {10.1103/PhysRevLett.116.061102},
  url = {https://link.aps.org/doi/10.1103/PhysRevLett.116.061102}
}

@article{ligo,
doi = {10.1088/0264-9381/32/7/074001},
url = {https://dx.doi.org/10.1088/0264-9381/32/7/074001},
year = {2015},
month = {mar},
publisher = {IOP Publishing},
volume = {32},
number = {7},
pages = {074001},
author = {J Aasi and others},
collaboration = {LIGO Scientific Collaboration},
title = "{Advanced LIGO}",
journal = {Classical and Quantum Gravity}
}

@article{virgo,
    author = "Acernese, F. and others",
    collaboration = "VIRGO",
    title = "{Advanced Virgo: a second-generation interferometric gravitational wave detector}",
    eprint = "1408.3978",
    archivePrefix = "arXiv",
    primaryClass = "gr-qc",
    doi = "10.1088/0264-9381/32/2/024001",
    journal = "Class. Quant. Grav.",
    volume = "32",
    number = "2",
    pages = "024001",
    year = "2015"
}

@article{gwtc-1,
  title = {{GWTC-1: A Gravitational-Wave Transient Catalog of Compact Binary Mergers Observed by LIGO and Virgo during the First and Second Observing Runs}},
  author = {Abbott, B. P. and others},
  collaboration = {LIGO Scientific Collaboration and Virgo Collaboration},
  journal = {Phys. Rev. X},
  volume = {9},
  issue = {3},
  pages = {031040},
  numpages = {49},
  year = {2019},
  month = {Sep},
  publisher = {American Physical Society},
  doi = {10.1103/PhysRevX.9.031040},
  url = {https://link.aps.org/doi/10.1103/PhysRevX.9.031040}
}

@misc{gwtc-2.1,
      title={{GWTC-2.1: Deep Extended Catalog of Compact Binary Coalescences Observed by LIGO and Virgo During the First Half of the Third Observing Run}}, 
      author={R. Abbott and others},
      collaboration = {LIGO Scientific Collaboration and Virgo Collaboration},
      year={2022},
      eprint={2108.01045},
      archivePrefix={arXiv},
      primaryClass={gr-qc},
      url={https://arxiv.org/abs/2108.01045}, 
}

@article{gwtc-3,
  title = {{GWTC-3: Compact Binary Coalescences Observed by LIGO and Virgo during the Second Part of the Third Observing Run}},
  author = {Abbott, R. and others},
  collaboration = {LIGO Scientific Collaboration, Virgo Collaboration, and KAGRA Collaboration},
  journal = {Phys. Rev. X},
  volume = {13},
  issue = {4},
  pages = {041039},
  numpages = {89},
  year = {2023},
  month = {Dec},
  publisher = {American Physical Society},
  doi = {10.1103/PhysRevX.13.041039},
  url = {https://link.aps.org/doi/10.1103/PhysRevX.13.041039}
}

@ARTICLE{gwtc-4,
       author = {{The LIGO Scientific Collaboration} and {the Virgo Collaboration} and {the KAGRA Collaboration}},
        title = "{GWTC-4.0: Updating the Gravitational-Wave Transient Catalog with Observations from the First Part of the Fourth LIGO-Virgo-KAGRA Observing Run}",
      journal = {arXiv e-prints},
         year = 2025,
        month = aug,
          eid = {arXiv:2508.18082},
        pages = {arXiv:2508.18082},
          doi = {10.48550/arXiv.2508.18082},
archivePrefix = {arXiv},
       eprint = {2508.18082},
 primaryClass = {gr-qc},
       adsurl = {https://ui.adsabs.harvard.edu/abs/2025arXiv250818082T}
}

@ARTICLE{gwtc-5,
       author = {{The LIGO Scientific Collaboration} and {the Virgo Collaboration} and {the KAGRA Collaboration}},
        title = "{GWTC-5.0: Observations from the Second Part of the Fourth LIGO-Virgo-KAGRA Observing Run and Updates to the Gravitational-Wave Transient Catalog}",
      journal = {arXiv e-prints},
         year = 2026,
        month = may,
          eid = {arXiv:2605.27225},
        pages = {arXiv:2605.27225},
          doi = {10.48550/arXiv.2605.27225},
archivePrefix = {arXiv},
       eprint = {2605.27225},
 primaryClass = {gr-qc},
       adsurl = {https://ui.adsabs.harvard.edu/abs/2026arXiv260527225T}
}

@misc{gracedb,
    author = {LIGO-Virgo-KAGRA Collaboration},
    title = {Gravitational-Wave Candidate Event Database},
    url={https://gracedb.ligo.org/superevents/public/}, 
     howpublished = {\url{https://gracedb.ligo.org/superevents/public/}}
}

@article{gw170817,
  title = "{GW170817: Observation of Gravitational Waves from a Binary Neutron Star Inspiral}",
  author = {Abbott, B. P. and others},
  collaboration = {LIGO Scientific Collaboration and Virgo Collaboration},
  journal = {Phys. Rev. Lett.},
  volume = {119},
  issue = {16},
  pages = {161101},
  numpages = {18},
  year = {2017},
  month = {Oct},
  publisher = {American Physical Society},
  doi = {10.1103/PhysRevLett.119.161101},
  url = {https://link.aps.org/doi/10.1103/PhysRevLett.119.161101}
}

@article{kagra,
    author = "Akutsu, T. and others",
    collaboration = "KAGRA",
    title = "{Overview of KAGRA: Detector design and construction history}",
    eprint = "2005.05574",
    archivePrefix = "arXiv",
    primaryClass = "physics.ins-det",
    doi = "10.1093/ptep/ptaa125",
    journal = "PTEP",
    volume = "2021",
    number = "5",
    pages = "05A101",
    year = "2021"
}

@misc{lisa,
      title={Laser Interferometer Space Antenna}, 
      author={Pau Amaro-Seoane and others},
      year={2017},
      eprint={1702.00786},
      archivePrefix={arXiv},
      primaryClass={astro-ph.IM},
      url={https://arxiv.org/abs/1702.00786}, 
}

@article{tsukada,
  title = "{Improved ranking statistics of the GstLAL inspiral search for compact binary coalescences}",
  author = {Tsukada, Leo and others},
  journal = {Phys. Rev. D},
  volume = {108},
  issue = {4},
  pages = {043004},
  numpages = {13},
  year = {2023},
  month = {Aug},
  publisher = {American Physical Society},
  doi = {10.1103/PhysRevD.108.043004},
  url = {https://link.aps.org/doi/10.1103/PhysRevD.108.043004}
}

@article{CANNON2021,
title = "{GstLAL: A software framework for gravitational wave discovery}",
journal = {SoftwareX},
volume = {14},
pages = {100680},
year = {2021},
issn = {2352-7110},
doi = {https://doi.org/10.1016/j.softx.2021.100680},
url = {https://www.sciencedirect.com/science/article/pii/S235271102100025X},
author = {Kipp Cannon and Sarah Caudill and Chiwai Chan and Bryce Cousins and Jolien D.E. Creighton and Becca Ewing and Heather Fong and Patrick Godwin and Chad Hanna and Shaun Hooper and Rachael Huxford and Ryan Magee and Duncan Meacher and Cody Messick and Soichiro Morisaki and Debnandini Mukherjee and Hiroaki Ohta and Alexander Pace and Stephen Privitera and Iris {de Ruiter} and Surabhi Sachdev and Leo Singer and Divya Singh and Ron Tapia and Leo Tsukada and Daichi Tsuna and Takuya Tsutsui and Koh Ueno and Aaron Viets and Leslie Wade and Madeline Wade}
}

@misc{gstreamer,
	howpublished = {\url{https://gstreamer.freedesktop.org/documentation/index.html}},
	title = {\uppercase{GS}treamer},
}

@misc{sgn,
    howpublished = {\url{https://git.ligo.org/greg/sgn}},
    title = {\uppercase{SGN}}
}

@misc{sgn-ts,
    howpublished = {\url{https://git.ligo.org/greg/sgn-ts}},
    title = {\uppercase{SGN-TS}}
}

@ARTICLE{sgnax,
       author = {{Yarbrough}, Zach and {Godwin}, Olivia and {Davis}, Derek and {Gonz{\'a}lez}, Gabriela},
        title = "{SGNAX: a unified matched-filter and excess-power pipeline for gravitational-wave detector characterization}",
      journal = {arXiv e-prints},
         year = 2026,
        month = aug,
          eid = {arXiv:2608.09804},
        pages = {arXiv:2608.09804},
          doi = {10.48550/arXiv.2608.09804},
archivePrefix = {arXiv},
       eprint = {2608.09804},
 primaryClass = {astro-ph.IM},
       adsurl = {https://ui.adsabs.harvard.edu/abs/2026arXiv260809804Y}
}

@misc{bottle,
    howpublished = {\url{https://bottlepy.org/docs/dev/}},
    title = {Bottle: Python Web Framework}
}

@misc{lalsuite,
       author         = "{LIGO Scientific Collaboration}",
       title          = "{LIGO} {A}lgorithm {L}ibrary - {LALS}uite",
       howpublished   = "free software (GPL)",
       doi            = "10.7935/GT1W-FZ16",
       year           = "2018"
}

@article{huang2025,
  title = "{Scalable matched-filtering pipeline for gravitational-wave searches of compact binary mergers}",
  author = {Huang, Yun-Jing and Hanna, Chad and Ewing, Becca and Godwin, Patrick and Gonsalves, Joshua and Magee, Ryan and Messick, Cody and Tsukada, Leo and Yarbrough, Zach and Joshi, Prathamesh and Kennington, James and Niu, Wanting and Rollins, Jameson and Shah, Urja},
  journal = {Phys. Rev. D},
  volume = {112},
  issue = {8},
  pages = {082002},
  numpages = {8},
  year = {2025},
  month = {Oct},
  publisher = {American Physical Society},
  doi = {10.1103/m7bv-c7rm},
  url = {https://link.aps.org/doi/10.1103/m7bv-c7rm}
}

@ARTICLE{indigo2,
       author = {{Unnikrishnan}, C.~S.},
        title = "{LIGO-India: A decadal assessment on its scope, relevance, progress and future}",
      journal = {International Journal of Modern Physics D},
         year = 2024,
        month = apr,
       volume = {33},
          eid = {2450025},
        pages = {2450025},
          doi = {10.1142/S0218271824500251},
       adsurl = {https://ui.adsabs.harvard.edu/abs/2024IJMPD..3350025U}
}

@ARTICLE{indigo1,
       author = {{Unnikrishnan}, C.~S.},
        title = "{IndIGO and Ligo-India Scope and Plans for Gravitational Wave Research and Precision Metrology in India}",
      journal = {International Journal of Modern Physics D},
         year = 2013,
        month = jan,
       volume = {22},
       number = {1},
          eid = {1341010},
        pages = {1341010},
          doi = {10.1142/S0218271813410101},
archivePrefix = {arXiv},
       eprint = {1510.06059},
 primaryClass = {physics.ins-det},
       adsurl = {https://ui.adsabs.harvard.edu/abs/2013IJMPD..2241010U}
}

@article{Huang2025_sgnl,
    author = "Huang, Yun-Jing and others",
    title = "{SGNL: Scalable Low-Latency Gravitational Wave Detection Pipeline for Compact Binary Mergers}",
    eprint = "2511.04730",
    archivePrefix = "arXiv",
    primaryClass = "astro-ph.IM",
    month = "11",
    year = "2025"
}

@ARTICLE{viets2018,
       author = {{Viets}, A.~D. and {Wade}, M. and {Urban}, A.~L. and {Kandhasamy}, S. and {Betzwieser}, J. and {Brown}, Duncan A. and {Burguet-Castell}, J. and {Cahillane}, C. and {Goetz}, E. and {Izumi}, K. and {Karki}, S. and {Kissel}, J.~S. and {Mendell}, G. and {Savage}, R.~L. and {Siemens}, X. and {Tuyenbayev}, D. and {Weinstein}, A.~J.},
        title = "{Reconstructing the calibrated strain signal in the Advanced LIGO detectors}",
      journal = {Classical and Quantum Gravity},
         year = 2018,
        month = may,
       volume = {35},
       number = {9},
          eid = {095015},
        pages = {095015},
          doi = {10.1088/1361-6382/aab658},
archivePrefix = {arXiv},
       eprint = {1710.09973},
 primaryClass = {astro-ph.IM},
       adsurl = {https://ui.adsabs.harvard.edu/abs/2018CQGra..35i5015V}
}

@article{godwin2020low,
  title={Low-latency statistical data quality in the era of multi-messenger astronomy},
  author={Godwin, Patrick},
  year={2020}
}

@article{Einstein:1916cc,
    author = "Einstein, Albert",
    title = "{Approximative Integration of the Field Equations of Gravitation}",
    journal = "Sitzungsber. Preuss. Akad. Wiss. Berlin (Math. Phys. )",
    volume = "1916",
    pages = "688--696",
    year = "1916"
}

@article{Einstein:1915ca,
    author = "Einstein, Albert",
    title = "{The Field Equations of Gravitation}",
    journal = "Sitzungsber. Preuss. Akad. Wiss. Berlin (Math. Phys. )",
    volume = "1915",
    pages = "844--847",
    year = "1915"
}

@ARTICLE{wade2025,
       author = {{Wade}, M. and {Betzwieser}, J. and {Bhattacharjee}, D. and {Dartez}, L. and {Goetz}, E. and {Kissel}, J. and {Sun}, L. and {Viets}, A. and {Carney}, M. and {Makelele}, E. and {Wade}, L.},
        title = "{Toward low-latency, high-fidelity calibration of the LIGO detectors with enhanced monitoring tools}",
      journal = {Classical and Quantum Gravity},
         year = 2025,
        month = nov,
       volume = {42},
       number = {21},
          eid = {215016},
        pages = {215016},
          doi = {10.1088/1361-6382/ae1095},
archivePrefix = {arXiv},
       eprint = {2508.08423},
 primaryClass = {gr-qc},
       adsurl = {https://ui.adsabs.harvard.edu/abs/2025CQGra..42u5016W}
}

@ARTICLE{wade2023,
       author = {{Wade}, M. and {Viets}, A.~D. and {Chmiel}, T. and {Stover}, M. and {Wade}, L.},
        title = "{Improving LIGO calibration accuracy by using time-dependent filters to compensate for temporal variations}",
      journal = {Classical and Quantum Gravity},
         year = 2023,
        month = feb,
       volume = {40},
       number = {3},
          eid = {035001},
        pages = {035001},
          doi = {10.1088/1361-6382/acabf6},
archivePrefix = {arXiv},
       eprint = {2207.00621},
 primaryClass = {astro-ph.IM},
       adsurl = {https://ui.adsabs.harvard.edu/abs/2023CQGra..40c5001W}
}








\end{document}